\documentclass[conference]{IEEEtran}
\IEEEoverridecommandlockouts
\usepackage[2023, cc-by, pages={1}{2}, url=https://osda.ws/r/SGTzU]{osda}
\usepackage{cite}
\usepackage{amsmath,amssymb,amsfonts}
\usepackage{algorithmic}
\usepackage{graphicx}
\usepackage{textcomp}
\usepackage{xcolor}
\def\BibTeX{{\rm B\kern-.05em{\sc i\kern-.025em b}\kern-.08em
    T\kern-.1667em\lower.7ex\hbox{E}\kern-.125emX}}

\begin{document}

\title{Hog 2023.1: a collaborative management tool to handle Git-based HDL repository}

\author{
  \IEEEauthorblockN{N.V. Biesuz\IEEEauthorrefmark{1},
    R.A. Rojas Caballero\IEEEauthorrefmark{2},
    D. Cieri\IEEEauthorrefmark{3},
    N. Giangiacomi\IEEEauthorrefmark{4},
    F. Gonnella\IEEEauthorrefmark{5},
    G. Loustau De Linares\IEEEauthorrefmark{2},
    A. Peck\IEEEauthorrefmark{6}}
  \IEEEauthorblockA{\IEEEauthorrefmark{1}INFN Sezione di Ferrara, Ferrara, Italy}
  \IEEEauthorblockA{\IEEEauthorrefmark{2}University of Massachusetts Amherst, Amherst, MA, United States}
  \IEEEauthorblockA{\IEEEauthorrefmark{3}Max-Planck-Institut f\"ur Physik, Munich, Germany}
  \IEEEauthorblockA{\IEEEauthorrefmark{4}Camlin Technologies Italy, Bologna, Italy}
  \IEEEauthorblockA{\IEEEauthorrefmark{5}University of Birmingham, Birmingham, United Kingdom}
  \IEEEauthorblockA{\IEEEauthorrefmark{6}Boston University, Boston, MA, United States}}

\maketitle

\begin{abstract}
  Hog (HDL on Git) is an open-source tool designed to manage Git-based HDL repositories. It aims to simplify HDL project development, maintenance, and versioning by using Git to guarantee synthesis and implementation reproducibility and binary file traceability. This is ensured by linking each produced binary file to a specific Git commit, embedding the Git commit hash (SHA) into the binary file via HDL generics stored in firmware registers.

  Hog is released twice a year, in January and in June. We present here the latest stable version 2023.1, which introduces major novel features, such as the support for Microchip Libero IDE, and the capability to run the Hog Continuous Integration (Hog-CI) workflow with GitHub Actions.

  A plan to integrate Hog with the OpenCores repository is also described, which is expected to be completed for Hog release 2023.2.
\end{abstract}


\section{Introduction}

HDL designs tend to be developed in an unorganized way, often leading to the duplication of significant portions of the code and the use of ad-hoc build scripts. Git is a good option to manage source code\cite{git}, but unfortunately the structure of HDL projects is often not well suited to be stored in a Git repository, with many files that are automatically modified by the HDL Integrated Development Environment (IDE) software.

Hog (HDL on Git) \cite{hog-repo, Hog:jinst, Hog:2023} aims to provide a solution to this problem which simplifies HDL project development, maintenance, and versioning, by exploiting Git as a comprehensive collaborative tool.

Hog guarantees the reproducibility of firmware synthesis, placement, and routing by maintaining absolute control over the source files, constraints files, and synthesis and implementation settings. IDE projects are reproducibly generated from the version controlled project configuration files. Furthermore, Hog embeds the firmware version and the Git hash numbers directly into the firmware binary file, by passing them as generics to the top module of the project before running the synthesis, to ensure the traceability of binary files, even in case they get renamed.

Hog is well integrated with Xilinx, Intel, and Microchip IDEs \cite{vivado,ise,quartus,libero} and allows the developers to use either the normal graphical interface or a command line workflow, which executes the software in batch mode.

Finally, Hog provides a documented set of YAML files, which can be used to easily set up Gitlab and Github Continuous Integration workflows. Hog-CI runs firmware implementation and simulation, and automatically creates tags and releases following the repository methodology described in previous publications \cite{Hog:jinst, Hog:2023}.

In the next sections, we describe the new features of the Hog 2023.1 release, namely:

\begin{itemize}
    \item \textbf{Libero support}. Support to Microchip Libero SoC HDL projects.
    \item \textbf{GitHub Actions}. Developers using GitHub will be able to run the Hog-CI using the GitHub Actions \cite{github-actions} framework.
    \item \textbf{Apptainer and Docker support}. Apptainer \cite{apptainer} or Docker images \cite{docker} can now be used to run the Hog-CI, simplifying the setup of the dedicated CI machines.
    \item \textbf{Multiple release branches}. Support for persistent release branches has been added, which allows developers patching older releases with hotfixes, and provides a second stage of control before releasing on the main branch of the repository.
    \item \textbf{Enhanced project configuration}. Custom user generics can now be defined and passed to the top level of the project, allowing for easy parameterization of projects.
\end{itemize}

\section{Hog support for Microchip Libero SoC}

The Hog team is planning to extend the support of Hog to all major FPGA development tools, and, in this context, support for Microchip Libero SoC has been added. Unfortunately, Libero SoC does not include a full version of the \texttt{tcllib} package, which is required by several functions in Hog. For this reason, developers must install \texttt{tcllib} on their machine, and set an environmental variable (\texttt{HOG\_TCLLIB\_PATH}) to tell Hog where to find the package.

Another difference compared to the standard Xilinx or Intel workflow is in the handling of IP cores. For Libero we do not provide support to core container files (e.g. \texttt{.xci} files in Vivado), so instead they must be generated by a Libero Tcl command which can be added to the Hog project using the \texttt{post-creation.tcl} file.

More information about the Libero support is available in the Hog documentation hub~\cite{hog-docs}. 

\section{Hog-CI with GitHub Actions}

Hog2023.1 provides templates to set up a Continuous Integration flow to be run within the GitHub Actions framework. We adopted the same methodology implemented for the GitLab CI/CD workflow \cite{gitlab-ci}.

The workflow uses both self-hosted and public runners. Any pull request towards the main (or develop/release, if enabled) branch of the GitHub repository activates an Actions pipeline, which builds and simulates the chosen projects. The CI is configurable by means of GitHub Actions secret and input variables, which allows, for example, enabling the generation of Doxygen documentation \cite{doxygen}.


Once the pull-request is accepted, a second pipeline is activated, which automatically tags the repository and optionally creates the GitHub release.

More details are provided in the Hog documentation hub~\cite{hog-docs}. 

\section{Hog-CI using Apptainer and Docker images}

To facilitate the setup of the machines for the Hog-CI, we introduced support for Apptainer and Docker images.

Two different approaches are employed. For Apptainer, users must copy the Apptainer image to an accessible path and then define an environmental variable (\texttt{HOG\_APPTAINER\_IMAGE}) which tells Hog to run the workflow with the provided Apptainer image.

To use Docker images, we took advantage of the good integration of Docker with GitHub and GitLab. Users only need to specify the location of the Docker image in the CI configuration YAML file, remembering to select private runners for execution. We highly recommend downloading the Docker container on the machines beforehand to speed up the loading time of the container.

\section{Multiple release branches}

The semantic versioning scheme ($M.m.p$) adopted by Hog considers only a single release branch, typically \texttt{master} or \texttt{main}, increasing by default the patch ($p$) number every time a feature branch is merged. 

We also introduced the possibility to have a second release branch for intermediate development. This allows developers to maintain a stable main branch while continuing to work on the develop branch. Hotfixes can be merged directly onto the main branch independently from the develop branch. In addition to this, multiple release branches can be kept on the repository, which allows developers to perform hotfixes on all past releases.

Both the versioning and the naming of the releases branch can be configured, as described in the documentation.

\section{Enhanced project configuration}

Often users require multiple projects sharing the same code, whose implementation is driven by different configurations. A new feature of Hog allows developers to set the values of generics/parameters directly in the \texttt{hog.conf} configuration file, which are then passed to the top level module of the project. This feature allows designs to be easily parameterised by simply having different versions of the project configuration file.

\section{Conclusions}

Hog 2023.1 was released in January 2023, and contains important new features. The Hog team is working on extending the compatibility of the tool to all major FPGA development software and plans to continue this work also in the future releases. We also intend to add the option to create standalone simulation projects, including support for open-source tools, such as GHDL \cite{ghdl} and cocotb \cite{cocotb}.

In the meantime, important developments have been made to improve the Hog continuous integration workflow, extending the support to GitHub actions. We plan to continue this work, keeping an eye on new features from GitLab or GitHub, which might be integrated in the workflow.

Hog is currently used by many projects within the upgrade of the trigger and data-acquisition systems of the ATLAS and CMS particle-physics experiments at CERN, in several astrophysics and synchrotron experiments, and also in some industries.

Finally, the Hog team is in contact with the OpenCores group \cite{opencores} for a future collaboration. A novel feature will allow users to directly add OpenCores designs as Git submodules in their projects. For this, some work will be done on the OpenCores repositories to make the modules more Hog friendly.

\end{document}